\documentclass[12pt,a4paper,english,aps,superscriptaddress,groupedaddress,showpacs,showkeys]{revtex4}

\usepackage[T1]{fontenc}
\usepackage[latin9]{inputenc}
\usepackage{babel}
\usepackage{amsmath}
\usepackage{amssymb}
\usepackage{esint}
\usepackage[unicode=true]
 {hyperref}
\usepackage{breakurl}

\makeatletter

\@ifundefined{textcolor}{}
{%
 \definecolor{BLACK}{gray}{0}
 \definecolor{WHITE}{gray}{1}
 \definecolor{RED}{rgb}{1,0,0}
 \definecolor{GREEN}{rgb}{0,1,0}
 \definecolor{BLUE}{rgb}{0,0,1}
 \definecolor{CYAN}{cmyk}{1,0,0,0}
 \definecolor{MAGENTA}{cmyk}{0,1,0,0}
 \definecolor{YELLOW}{cmyk}{0,0,1,0}
}

\usepackage{dcolumn}
\usepackage{bm}
\usepackage{amsfonts}

\makeatother

\begin{document}

\title{Characterizing Klein-Fock-Gordon-Majorana particles in (1+1) dimensions}

\author{Salvatore De Vincenzo}

\homepage{https://orcid.org/0000-0002-5009-053X}

\email{[salvatored@nu.ac.th]}

\selectlanguage{english}%

\affiliation{The Institute for Fundamental Study (IF), Naresuan University, Phitsanulok
65000, Thailand}

\date{November 27, 2023}

\begin{abstract}
\noindent \textbf{Abstract} Theoretically, in (1+1) dimensions, one
can have Klein-Fock-Gordon-Majorana (KFGM) particles. More precisely,
these are one-dimensional (1D) Klein-Fock-Gordon (KFG) and Majorana
particles at the same time. In principle, the wave equations considered
to describe such first-quantized particles are the standard 1D KFG
equation and/or the 1D Feshbach-Villars (FV) equation, each with a
real Lorentz scalar potential and some kind of Majorana condition.
The aim of this paper is to analyze the latter assumption fully and
systematically; additionally, we introduce specific equations and
boundary conditions to characterize these particles when they lie
within an interval (or on a line with a tiny hole at a point). In
fact, we write first-order equations in the time derivative that do
not have a Hamiltonian form. We may refer to these equations as first-order
1D Majorana equations for 1D KFGM particles. Moreover, each of them
leads to a second-order equation in time that becomes the standard
1D KFG equation when the scalar potential is independent of time.
Additionally, we examine the nonrelativistic limit of one of the first-order
1D Majorana equations. 
\end{abstract}



\maketitle

\section{Introduction}

\noindent In (3+1) dimensions, there is the possibility that a spin-$0$
particle is its own antiparticle. A typical example of this is the
neutral pion (or neutral pi meson) $\Pi^{0}$ (although it is not
exactly an elementary particle) \cite{RefA,RefB}. We may refer to
these particles as three-dimensional (3D) Klein-Fock-Gordon-Majorana
(KFGM) particles. We recall that, in general, a Majorana particle
is its own antiparticle, i.e., it is a strictly neutral particle \cite{RefC},
and the wavefunction that characterizes it is invariant under the
respective charge-conjugation operation \cite{RefD,RefE,RefF} (this
is, in principle, within a phase factor). The latter specific fact
is what defines a Majorana particle and is called the Majorana condition.
Among the known spin-$\tfrac{1}{2}$ particles, only neutrinos could
be of a Majorana nature, i.e., only neutrinos could be Majorana fermions
\cite{RefG}. Similarly, because photons (spin-$1$) and gravitons
(spin-$2$) are also strictly neutral particles, we may say that they
are also of a Majorana nature \cite{RefG,RefH}. Incidentally, completely
different types of Majorana particles can even be found in certain
condensed-matter systems described in the second quantization formalism.
These particles emerge as quasiparticles excitations that are their
own antiparticles and whose statistics is not fermionic \cite{RefG,RefI,RefJ,RefK,RefL}.
Incidentally, these excitations have been called Majorinos \cite{RefM,RefN}.

Thus, in the first quantization, we may say that the primordial wave
equation intended to describe a strictly neutral 3D KFG spin-$0$
particle, i.e., a 3D KFGM spin-$0$ particle, is the 3D Klein-Fock-Gordon
(KFG) equation in its standard form with a real-valued Lorentz scalar
interaction (or potential) \cite{RefA,RefB,RefO}, but in addition,
together with some Majorana conditions. Likewise, the 3D Feshbach-Villars
(FV) equation with the scalar potential and the respective Majorana
condition may also be used \cite{RefP}. Naturally, this way of characterizing
a 3D KFGM particle can also be implemented to describe a 1D KFGM particle
(the latter is a KFGM particle living in one space and one time dimension).
Thus, in this case, we may also use the 1D KFG equation in its standard
form and/or the 1D FV equation, both with a real-valued scalar potential
together with their respective Majorana condition.

The subject of neutral 3D KFG particles in the first quantization
(as well as in the second quantization) is generally mentioned in
books on relativistic quantum mechanics \cite{RefA,RefB,RefC,RefQ}.
Additionally, a neutral 3D KFG particle may not be equal to its antiparticle,
for example, a neutral $\mathrm{K}^{0}$ meson (or neutral kaon) is
different from its antiparticle $\overline{\mathrm{K}}^{0}$. In this
case, these two particles carry different internal attributes (in
fact, different hypercharges) and can be described by (classical)
complex fields or complex solutions of the standard 3D KFG equation
\cite{RefA,RefR}. However, if a neutral 3D KFG particle is equal
to its antiparticle, then there are no internal attributes that distinguish
them; consequently, they must be described by (classical) real fields
or real solutions of the standard 3D KFG equation, as usual \cite{RefA,RefC,RefG}.
Incidentally, we have also seen some references that roughly question
the use of real solutions to describe a strictly neutral 3D KFG particle
(see, for example, Refs. \cite{RefQ,RefS,RefT,RefU}). 

Actually, there is a well-known connection between the complexity
of any solution of the standard 3D KFG equation (which is a Lorentz
scalar) and the internal attribute electric charge. That is, if a
solution, i.e., $\psi$, describes the particle, then its complex
conjugate $\psi^{*}$ describes the antiparticle \cite{RefR}. In
fact, the usual densities $\varrho=\varrho(\mathrm{\mathbf{r}},t)=(\mathrm{i}\hbar/2\mathrm{m}c^{2})(\psi^{*}\,\partial_{t}\psi-\psi\,\partial_{t}\psi^{*})-(V/\mathrm{m}c^{2})\psi^{*}\psi$
and $\mathrm{\mathbf{j}}=\mathrm{\mathbf{j}}(\mathrm{\mathbf{r}},t)=-(\mathrm{i}\hbar/2\mathrm{m})(\psi^{*}\,\nabla\psi-\psi\,\nabla\psi^{*})$,
which satisfy a continuity equation (i.e., $\partial_{t}\varrho+\nabla\cdot\mathrm{\mathbf{j}=0}$),
change sign when the replacements $\psi\rightarrow\psi^{*}$, $\psi^{*}\rightarrow\psi$
and $V\rightarrow-V$ are made ($V$ is a real potential). Incidentally,
the latter result is also valid when there is additionally a (real)
Lorentz scalar potential (because $\varrho$ and $\mathrm{\mathbf{j}}$
do not depend on this type of potential). Consequently, it is clear
that for real solutions, $\psi=\psi^{*}$ (with $V=0$), $\varrho$
and $\mathrm{\mathbf{j}}$ vanish, i.e., there is no place for a conserved
current density four-vector for the strictly neutral 3D KFG particle
\cite{RefC,RefQ}. This conclusion seems to be a general property
of other strictly neutral (bosonic) particles \cite{RefC}.

The following is the path that our study follows. We consider the
(1+1)-dimensional case only. In the first part of Section II, we begin
by discussing the conditions that must be imposed on the electric
and scalar potentials such that the one-component solutions $\psi$
of the standard 1D KFG wave equation can be written as real solutions.
We also analyze the existence of complex solutions for this equation,
i.e., complex but not pure imaginary solutions. Then, we introduce
the 1D KFG wave equation in Hamiltonian form, i.e., the 1D FV wave
equation (its solutions $\Psi$ are two-component wavefunctions and
the Hamiltonian operator $\mathrm{\hat{h}}$ is a $2\times2$ matrix),
and again, we only include the electric potential and the Lorentz
scalar potential. We show that this equation cannot have real solutions
regardless of the real or complex nature of the potentials. We also
introduce the charge-conjugate wavefunction and the respective Hamiltonian
operator for this wavefunction. We show that if $\Psi$ describes
a 1D KFG particle in the presence of the potentials $V$ and $S$,
then $\Psi_{c}$ (its charge-conjugate wavefunction) describes a 1D
KFG particle in the potentials $-V^{*}$ and $S^{*}$. We then introduce
a first Majorana condition that defines a 1D KFGM particle, namely,
$\Psi=\Psi_{c}$, which suggests that the two components of $\Psi$
are no longer independent. Furthermore, this condition also implies
that $V$ must be a purely imaginary potential, i.e., $V=-V^{*}$
(consequently, $V$ must be zero if it can be considered real), and
$S$ must be real. We also obtain the reality condition $\psi=\psi^{*}$
as a consequence of imposing this Majorana condition. A second Majorana
condition that defines a 1D KFGM particle is given by $\Psi=-\Psi_{c}$,
and it imposes the same restrictions on the potentials as the first
Majorana condition; however, this time, we have that $\psi$ satisfies
the relation $\psi=-\psi^{*}$, i.e., $\psi$ is purely imaginary
but can obviously be written as real by writing $(\psi-\psi^{*})/2\mathrm{i}=\psi/\mathrm{i}$.
Thus, due to the Majorana conditions, the components of the wavefunction
$\Psi$ are always related to each other, which allows us to write
equations for only one component and to obtain the other component
algebraically. However, these equations are not of the Hamiltonian
type, i.e., they do not have the form $(\mathrm{i}\hbar\,\partial_{t}-\hat{\mathrm{H}})\phi=0$. 

In the second part of Section II, we show that the further imposition
of the formal pseudohermiticity condition on the Hamiltonian $\hat{\mathrm{h}}$
implies that the electric potential satisfies the relation $V=V^{*}$.
The latter formula together with the condition $V=-V^{*}$ gives us
$V=0$. In fact, if we place a 1D KFGM particle in the interior of
an interval, for example, $\Omega=[a,b]$, the operator $\mathrm{\hat{h}}$
with $V=0$ and a scalar potential $S\in\mathbb{R}$ is a pseudo self-adjoint
operator. We show that as a consequence of the latter property, the
respective solutions $\Psi$ of the 1D FV wave equation must satisfy
any boundary condition that is included in a general set of boundary
conditions that depends on three real parameters. Similarly, the respective
solutions $\psi$ of the standard 1D KFG equation must satisfy any
boundary condition that is included in its own real three-parameter
general family of boundary conditions. This is because the solutions
$\psi$ must be written as real if they are to describe a 1D KFGM
particle. It is worth mentioning that these general sets of boundary
conditions are the same for both types of Majorana conditions and
the most general for a 1D KFGM particle that is within an interval.
In fact, the most general sets of boundary conditions for a 1D KFG
particle in an interval were obtained in Ref. \cite{RefV}, and the
general sets of boundary conditions for a 1D KFGM particle presented
here arise precisely from those. Here, we obtain the most general
sets of boundary conditions for each component of the FV wavefunction
(these results depend on the type of Majorana condition one chooses).
Certainly, each component satisfies its own differential equation,
i.e., its own first-order 1D Majorana equation in the time derivative
(which we also present here). Additionally, the (complex) solutions
of these equations have the following characteristic: If the Majorana
condition is given by $\Psi=\Psi_{c}$ ($\Psi=-\Psi_{c}$), then the
real (imaginary) parts of the solutions satisfy the real second-order
1D KFG equation in time, and the imaginary (real) parts are simply
the time derivatives of the real (imaginary) parts. In Section III,
we present a summary and our conclusions. Additionally, in the Appendix
A, we show that each component of the 1D FV equation also satisfies
its own second-order 1D Majorana differential equation in time; however,
these equations become the standard 1D KFG equation when the scalar
potential is independent of time. Finally, in the Appendix B, we examine
the nonrelativistic limit of one of the first-order 1D Majorana equations. 

\section{Characterizing a 1D KFGM particle}
\subsection{Preliminaries}

\noindent Let us begin by writing the 1D KFG wave equation in its
standard form or the second order in time KFG equation in one spatial
dimension,
\begin{equation}
\left[(\hat{\mathrm{E}}-V)^{2}-(c\,\hat{\mathrm{p}})^{2}-(\mathrm{m}c^{2})^{2}-2\,\mathrm{m}c^{2}S\,\right]\psi=0,
\end{equation}
where $\psi=\psi(x,t)$ is a one-component wavefunction, $\hat{\mathrm{E}}=\mathrm{i}\hbar\,\partial/\partial t$
is the energy operator, $\hat{\mathrm{p}}=-\mathrm{i}\hbar\,\partial/\partial x$
is the momentum operator, $V=V(x)$ is the electric potential (energy),
and $S=S(x,t)\in\mathbb{R}$ is a real-valued Lorentz scalar potential
(energy). Thus, the latter equation can also be written as follows:
\begin{equation}
\left[\,-\hbar^{2}\frac{\partial^{2}}{\partial t^{2}}-2V\mathrm{i}\hbar\frac{\partial}{\partial t}+V^{2}+\hbar^{2}c^{2}\frac{\partial^{2}}{\partial x^{2}}-(\mathrm{m}c^{2})^{2}-2\,\mathrm{m}c^{2}S\,\right]\psi=0.
\end{equation}
Clearly, the differential operator acting on $\psi$ in Eq. (2) is
real when the potential $V$ is purely imaginary (this is somewhat
trivial but unexpected). Additionally, this operator is real when
$V=0$, i.e., when there is only a (real) scalar potential (throughout
the article, we consider $S$ to be a real potential, unless explicitly
stated otherwise). Consequently, in these two cases, the solutions
of the time-dependent 1D KFG equation in (2) can always be chosen
to be real, i.e., this equation can have solutions \textit{\`a la} Majorana.
In these two cases, the solutions of Eq. (2) need not be real, i.e.,
complex solutions can also be written (naturally, we refer to complex
solutions, i.e., we are not thinking of pure imaginary solutions,
unless that is explicitly stated). In fact, if $\hat{\mathrm{L}}$
is a real operator, then $\psi$ and $\psi^{*}$ are solutions of
the differential equation $\hat{\mathrm{L}}\psi=0$, and its real
solutions are given by $(\psi+\psi^{*})/2$ and $(\psi-\psi^{*})/2\mathrm{i}$
(the superscript $^{*}$ denotes the complex conjugate). Certainly,
it is also important to note that whether the electric potential $V\in\mathbb{R}$
is not zero and $S\in\mathbb{R}$ exists (or not), the differential
operator acting on $\psi$ in Eq. (2) is complex, and therefore, the
solutions of the time-dependent 1D KFG equation in (2) are necessarily
complex. 

Now, let us introduce the following functions \cite{RefP}: 
\begin{equation}
\varphi+\chi=\psi
\end{equation}
and 
\begin{equation}
\varphi-\chi=\frac{1}{\mathrm{m}c^{2}}(\hat{\mathrm{E}}-V)\psi.
\end{equation}
By using Eqs. (3), (4) and (1), we obtain a system of coupled differential
equations for $\varphi$ and $\chi$, namely,
\begin{equation}
\hat{\mathrm{E}}\,\varphi=\left(\frac{\hat{\mathrm{p}}^{2}}{2\mathrm{m}}+S\right)(\varphi+\chi)+(\mathrm{m}c^{2}+V)\varphi,
\end{equation}
\begin{equation}
\hat{\mathrm{E}}\,\chi=-\left(\frac{\hat{\mathrm{p}}^{2}}{2\mathrm{m}}+S\right)(\varphi+\chi)-(\mathrm{m}c^{2}-V)\chi.
\end{equation}
The latter system can be written in matrix form, namely, 
\begin{equation}
\hat{\mathrm{E}}\,\hat{1}_{2}\Psi=\mathrm{\hat{h}}\Psi,
\end{equation}
where 
\begin{equation}
\hat{\mathrm{h}}=\frac{\hat{\mathrm{p}}^{2}}{2\mathrm{m}}(\hat{\tau}_{3}+\mathrm{i}\hat{\tau}_{2})+\mathrm{m}c^{2}\hat{\tau}_{3}+V\,\hat{1}_{2}+S\,(\hat{\tau}_{3}+\mathrm{i}\hat{\tau}_{2})
\end{equation}
is the Hamiltonian operator, $\Psi=\Psi(x,t)=\left[\,\varphi\;\,\chi\,\right]^{\mathrm{T}}=\left[\,\varphi(x,t)\;\,\chi(x,t)\,\right]^{\mathrm{T}}$
is the two-component column state vector (the symbol $^{\mathrm{T}}$
represents the transpose of a matrix), $\hat{\tau}_{3}=\hat{\sigma}_{z}$
and $\hat{\tau}_{2}=\hat{\sigma}_{y}$ are Pauli matrices, and $\hat{1}_{2}$
is the $2\times2$ identity matrix. The equation in (7), with $\hat{\mathrm{h}}$
given in Eq. (8), is the 1D FV wave equation with an electric potential
and a scalar potential (and is also called the 1D KFG equation in
Hamiltonian form) \cite{RefP}. Although this equation has not received
sufficient attention when addressing problems within the 1D KFG theory
in (1+1) dimensions, a few papers in which it is considered can be
found in Refs. \cite{RefW,RefX,RefY}. By using Eqs. (3) and (4),
we can explicitly write the relation between the one-component wavefunction
$\psi$ and the two-component column state vector (or wavefunction)
$\Psi$, namely,
\begin{equation}
\Psi=\left[\begin{array}{c}
\varphi\\
\chi
\end{array}\right]=\frac{1}{2}\left[\begin{array}{c}
\psi+\frac{1}{\mathrm{m}c^{2}}(\hat{\mathrm{E}}-V)\psi\\
\psi-\frac{1}{\mathrm{m}c^{2}}(\hat{\mathrm{E}}-V)\psi
\end{array}\right].
\end{equation}
Certainly, in this last expression, the scalar potential does not
appear. We note that even if $\psi$ is a real function, the components
of $\Psi$, $\varphi$ and $\chi$, are always complex, i.e., $\Psi$
is inexorably complex. In fact, if one has a wave equation of the
form $(\hat{\mathrm{E}}-\hat{\mathrm{H}})\Phi=0$, then one also has
that $(\hat{\mathrm{E}}-\hat{\mathrm{H}})^{*}\Phi=0$ if $\mathrm{i}\hat{\mathrm{H}}=(\mathrm{i}\hat{\mathrm{H}})^{*}$,
i.e., if $\mathrm{i}\hat{\mathrm{H}}$ is a real operator. This result
tells us that the time-dependent 1D FV wave equation cannot have real
solutions. Clearly, the same goes for the time-dependent Schr\"{o}dinger
equation. That is, these time-dependent wave equations cannot have
real solutions. 

The charge conjugate of $\Psi$, 
\begin{equation}
\Psi_{c}\equiv\hat{\tau}_{1}\Psi^{*},
\end{equation}
where $\hat{\tau}_{1}=\hat{\sigma}_{x}$ is a Pauli matrix, satisfies
the following equation: 
\begin{equation}
\hat{\mathrm{E}}\,\hat{1}_{2}\Psi_{c}=\mathrm{\hat{h}}_{c}\Psi_{c},
\end{equation}
where the respective Hamiltonian operator is given by 
\begin{equation}
\hat{\mathrm{h}}_{c}\equiv\frac{\hat{\mathrm{p}}^{2}}{2\mathrm{m}}(\hat{\tau}_{3}+\mathrm{i}\hat{\tau}_{2})+\mathrm{m}c^{2}\hat{\tau}_{3}+V_{c}\,\hat{1}_{2}+S_{c}\,(\hat{\tau}_{3}+\mathrm{i}\hat{\tau}_{2}).
\end{equation}
Taking the complex conjugate of Eq. (7), with $\hat{\mathrm{h}}$
given in Eq. (8), also using the results $\hat{\mathrm{E}}^{*}=-\hat{\mathrm{E}}$
and $(\hat{\mathrm{p}}^{2})^{*}=\hat{\mathrm{p}}^{2}$, and the facts
that $(\hat{\tau}_{3}+\mathrm{i}\hat{\tau}_{2})^{*}=(\hat{\tau}_{3}+\mathrm{i}\hat{\tau}_{2})$,
$(\hat{\tau}_{3})^{*}=\hat{\tau}_{3}$, and $\hat{\tau}_{1}\hat{\tau}_{3}=-\hat{\tau}_{3}\hat{\tau}_{1}$,
$\hat{\tau}_{1}\hat{\tau}_{2}=-\hat{\tau}_{2}\hat{\tau}_{1}$ ($\Rightarrow(\hat{\tau}_{3}+\mathrm{i}\hat{\tau}_{2})\hat{\tau}_{1}=-\hat{\tau}_{1}(\hat{\tau}_{3}+\mathrm{i}\hat{\tau}_{2})$),
and $\hat{\tau}_{1}^{2}=\hat{1}_{2}$, and finally, using the definition
of $\Psi_{c}$ in Eq. (10), we obtain the same Eq. (11), but $\hat{\mathrm{h}}_{c}$
is given by
\begin{equation}
\hat{\mathrm{h}}_{c}=\frac{\hat{\mathrm{p}}^{2}}{2\mathrm{m}}(\hat{\tau}_{3}+\mathrm{i}\hat{\tau}_{2})+\mathrm{m}c^{2}\hat{\tau}_{3}-V^{*}\,\hat{1}_{2}+S\,(\hat{\tau}_{3}+\mathrm{i}\hat{\tau}_{2}).
\end{equation}
Comparing the latter operator with the operator given in Eq. (12),
we obtain the following two relations: 
\begin{equation}
V_{c}=-V^{*}\:,\quad S_{c}=S.
\end{equation}
If we had considered placing a complex scalar potential in the Hamiltonian
$\hat{\mathrm{h}}$ given in Eq. (8), then the Hamiltonian $\hat{\mathrm{h}}_{c}$
would be the one given in Eq. (13) but with the replacement $S\rightarrow S^{*}$,
and therefore, the corresponding relation in Eq. (14) would be $S_{c}=S^{*}$. 

Equation (7) describes via $\Psi$ a 1D KFG particle in the presence
of the potentials $V$ and $S$. Likewise, Eq. (11) describes via
$\Psi_{c}$ a 1D KFG particle in the presence of the potentials $-V^{*}$
and $S$. For example, if one sets $V\in\mathbb{R}$ and $S\in\mathbb{R}$,
then one has that $\hat{\mathrm{E}}\,\hat{1}_{2}\Psi=\mathrm{\hat{h}}(V)\Psi$
(see Eq. (8)) and $\hat{\mathrm{E}}\,\hat{1}_{2}\Psi_{c}=\mathrm{\hat{h}}_{c}(V)\Psi_{c}=\mathrm{\hat{h}}(-V)\Psi_{c}$
(see Eq. (13)), i.e., $\Psi$ describes a 1D KFG particle with one
sign of electric charge, and $\Psi_{c}$ describes the 1D KFG particle
with the opposite sign of electric charge (i.e., its antiparticle).

Now, let us explore the possibility that a 1D KFG particle is its
own antiparticle; therefore, it must be an electrically and strictly
neutral particle. The condition that defines a particle of this type
is customarily given by
\begin{equation}
\Psi=\Psi_{c}.
\end{equation}
We refer to this relation as the standard Majorana condition. The
latter condition imposes the following relation between the components
of $\Psi$: $\varphi=\varphi_{c}=\chi^{*}$ ($\Leftrightarrow\chi=\chi_{c}=\varphi^{*}$).
Hence, $\Psi=\left[\,\chi^{*}\;\,\chi\,\right]^{\mathrm{T}}$ ($\Leftrightarrow\Psi=\left[\,\varphi\;\,\varphi^{*}\,\right]^{\mathrm{T}}$),
i.e., $\varphi$ and $\chi$ are not independent. Then, comparing
the 1D FV wave equations for $\Psi$ (see Eqs. (7) and (8)) and $\Psi_{c}\:(=\Psi)$
(see Eqs. (11) and (12)), and by using the relations in Eq. (14),
we obtain 
\begin{equation}
V=-V^{*}.
\end{equation}
That is, the complex potential $V$ must be purely imaginary, but
if $V$ had been chosen as real, then $V$ must be zero (because $V=-V$).
We already know that the potential $S$ is real valued, but if from
the beginning we had decided to place a complex potential $S$, then,
in addition to the relation given in Eq. (16), we can obtain $S=S^{*}$
(because $S_{c}=S^{*}$ and we also used the Majorana condition).
That is, the Majorana condition imposes that $S$ be a real scalar
potential (see the comment following Eq. (14)). Thus, in principle,
the 1D FV wave equation describing a 1D KFG particle that is also
a one-dimensional Majorana particle, i.e., a 1D KFGM particle, can
be written as follows: 
\begin{equation}
\hat{\mathrm{E}}\,\hat{1}_{2}\Psi=\mathrm{\hat{h}}\Psi=\left[\,\frac{\hat{\mathrm{p}}^{2}}{2\mathrm{m}}(\hat{\tau}_{3}+\mathrm{i}\hat{\tau}_{2})+\mathrm{m}c^{2}\hat{\tau}_{3}+V\,\hat{1}_{2}+S\,(\hat{\tau}_{3}+\mathrm{i}\hat{\tau}_{2})\,\right]\Psi,
\end{equation}
where if $V\in\mathbb{C}$, then it must be imaginary; and if $V\in\mathbb{\mathbb{R}}$,
then it must be zero. Likewise, the Lorentz scalar potential $S$
must be real. Additionally, the wavefunction $\Psi$ must have the
form $\Psi=\left[\,\chi^{*}\;\,\chi\,\right]^{\mathrm{T}}$ or $\Psi=\left[\,\varphi\;\,\varphi^{*}\,\right]^{\mathrm{T}}$. 

Equivalently, it should be noted that the 1D FV wave equation given
in Eqs. (7) and (8), or better, the coupled system of equations given
in Eqs. (5) and (6), is invariant under the following substitution:
\begin{equation}
\Psi=\left[\,\varphi\;\,\chi\,\right]^{\mathrm{T}}\:\rightarrow\:\Psi_{c}=\left[\,\chi^{*}\;\,\varphi^{*}\,\right]^{\mathrm{T}},
\end{equation}
but the conditions $V=-V^{*}$ and $S=S^{*}$ must be satisfied. In
other words, if the latter conditions are satisfied, then $\Psi$
and $\Psi_{c}$ satisfy the same equation (the latter is the equation
for the 1D KFGM particle, namely, Eq. (17)). 

Now, let us study the consequences of imposing the Majorana condition
given in Eq. (15) on the two-component column vector $\Psi$ given
in Eq. (9). Because $\hat{\mathrm{E}}^{*}=-\hat{\mathrm{E}}$, and
$V^{*}=-V$ for the 1D KFGM particle (see Eq. (16)), it follows that
\[
\Psi=\frac{1}{2}\left[\begin{array}{c}
\psi+\frac{1}{\mathrm{m}c^{2}}(\hat{\mathrm{E}}-V)\psi\\
\psi-\frac{1}{\mathrm{m}c^{2}}(\hat{\mathrm{E}}-V)\psi
\end{array}\right]=\left[\begin{array}{cc}
0 & 1\\
1 & 0
\end{array}\right]\,\frac{1}{2}\left[\begin{array}{c}
\psi^{*}-\frac{1}{\mathrm{m}c^{2}}(\hat{\mathrm{E}}-V)\psi^{*}\\
\psi^{*}+\frac{1}{\mathrm{m}c^{2}}(\hat{\mathrm{E}}-V)\psi^{*}
\end{array}\right]=\hat{\tau}_{1}\Psi^{*}=\Psi_{c},
\]
from which one immediately obtains the result
\begin{equation}
\psi=\psi^{*},
\end{equation}
which is the reality condition for the wavefunction $\psi$ and can
be considered the standard Majorana condition for these solutions,
i.e., $\psi=\psi^{*}\equiv\psi_{c}$, where $\psi_{c}$ is the charge
conjugate of $\psi$ \cite{RefZ}. The latter relation also arises
immediately when using Eq. (3) and the Majorana condition in terms
of the components of $\Psi$ (i.e., $\varphi=\chi^{*}$), namely,
$\psi=\varphi+\chi=\chi^{*}+\varphi^{*}=(\chi+\varphi)^{*}=\psi^{*}$.
Thus, if the solutions of the standard 1D KGF wave equation given
in Eq. (2) describe a 1D KFGM particle; then, they must be real. However,
$V$ must be an imaginary potential (see Eq. (16)), or zero (if $V\in\mathbb{R}$),
and $S$ must be a real potential. Consistently, when the latter conditions
on the potentials are imposed on the operator acting on $\psi$ in
Eq. (2), the operator is real, and the solutions $\psi$ can be written
real. Obviously, the solutions of the time-dependent 1D FV wave equation
(17), $\Psi$, do not have to be real to describe a 1D KFGM particle;
as we know, these solutions are not real even when $\psi=\psi^{*}$
(see the comment that follows Eq. (9)). The latter is a situation
somewhat similar to that which occurs in Dirac theory in (1+1) dimensions.
Indeed, only in the Majorana representation can the solutions of the
Dirac equation describing a 1D Dirac-Majorana particle be real valued,
but in any other representation, the solutions of the Dirac equation
describing this particle are complex valued. Certainly, all these
solutions satisfy the Majorana condition \cite{ReZZ}.

Similarly, the 1D FV wave equation is invariant under the substitution
\begin{equation}
\Psi=\left[\,\varphi\;\,\chi\,\right]^{\mathrm{T}}\:\rightarrow\:-\Psi_{c}=\left[\,-\chi^{*}\;\,-\varphi^{*}\,\right]^{\mathrm{T}},
\end{equation}
and again, the conditions $V=-V^{*}$ and $S=S^{*}$ must be satisfied.
That is, $\Psi$ and $-\Psi_{c}$ satisfy the same equation, but this
time, we obtain the result $\psi=-\psi^{*}$ (to prove this, one can
use the same procedure that led us to Eq. (19)). Thus, in this case,
the solutions $\psi$ are imaginary, but they can be written as real
simply by writing $(\psi-\psi^{*})/2\mathrm{i}=\psi/\mathrm{i}$.
In conclusion, the Majorana condition, which relates the two-component
wavefunction $\Psi$ to its charge-conjugate state $\Psi_{c}$ (or
$\Psi_{c}$ to its charge-conjugate state $\Psi$), appears here in
two forms, one standard form, $\Psi=\Psi_{c}$, and, say, one nonstandard
form, $\Psi=-\Psi_{c}$. In both cases, the one-component solution
$\psi$ can (and must) be written real, but additionally, the potentials
must satisfy the conditions $V=-V^{*}$ and $S=S^{*}$. Thus, the
Majorana condition $\Psi=-\Psi_{c}$ imposes the following relation
between the components of $\Psi$: $\varphi=-\varphi_{c}=-\chi^{*}$
($\Leftrightarrow\chi=-\chi_{c}=-\varphi^{*}$); hence, $\Psi=\left[\,-\chi^{*}\;\,\chi\,\right]^{\mathrm{T}}$
($\Leftrightarrow\Psi=\left[\,\varphi\;\,-\varphi^{*}\,\right]^{\mathrm{T}}$).
Using the relation $\varphi=-\chi^{*}$ and Eq. (3), the condition
$\psi=-\psi^{*}$ can also be obtained, namely, $\psi=\varphi+\chi=-\chi^{*}-\varphi^{*}=-(\chi+\varphi)^{*}=-\psi^{*}$.
In principle, the existence of two Majorana conditions defines two
specific and different types of 1D KFGM particles (this is the case
for 3D KFGM particles). When $\Psi=\Psi_{c}$, we have $+\Psi=\Psi_{c}=\hat{\tau}_{1}\Psi^{*}\equiv\hat{\mathrm{C}}\Psi$,
i.e., $\hat{\mathrm{C}}\Psi=+\Psi$, and when $-\Psi=\Psi_{c}$, we
have $-\Psi=\Psi_{c}=\hat{\tau}_{1}\Psi^{*}\equiv\hat{\mathrm{C}}\Psi$,
i.e., $\hat{\mathrm{C}}\Psi=-\Psi$. Then, for one of these particles,
its respective wavefunction is an eigenfunction of the charge conjugation
transformation (or operator) $\hat{\mathrm{C}}$ with the eigenvalue
$+1$, and for the other particle, its wavefunction is an eigenfunction
of $\hat{\mathrm{C}}$ with the eigenvalue $-1$ \cite{RefB,RefP}.
For example, the wavefunction corresponding to the 3D KFGM neutral
pion $\Pi^{0}$ is an eigenfunction of $\hat{\mathrm{C}}$ with eigenvalue
$+1$, i.e., the so-called C-parity of this particle is $+1$ \cite{RefP}. 

Thus far, the wave equation that we have considered to describe a
1D KFGM particle is Eq. (17). Because the components $\varphi$ and
$\chi$ of $\Psi$ are not independent, one can write an equation
for only one of these components and can obtain the other component
algebraically. In effect, Eq. (17) is the system of equations given
in Eqs. (5) and (6) with $V=-V^{*}$ and $S=S^{*}$. Then, if we take
Eq. (5) and use the Majorana condition $\Psi=\Psi_{c}$, namely, $\chi=\varphi^{*}$,
we obtain the following wave equation for the 1D KFGM particle:
\begin{equation}
\hat{\mathrm{E}}\,\varphi=\left(\frac{\hat{\mathrm{p}}^{2}}{2\mathrm{m}}+S\right)(\varphi+\varphi^{*})+(\mathrm{m}c^{2}+V)\varphi.
\end{equation}
From the solution $\varphi$ of the latter equation, the two-component
wavefunction $\Psi$ can be written immediately, namely, $\Psi=\left[\,\varphi\;\,\chi\,\right]^{\mathrm{T}}=\left[\,\varphi\;\,\varphi^{*}\,\right]^{\mathrm{T}}=\Psi_{c}$.
Alternatively, if we take Eq. (6) and use the Majorana condition $\varphi=\chi^{*}$,
we obtain an equation for the one-component wavefunction $\chi$,
namely, 
\begin{equation}
\hat{\mathrm{E}}\,\chi=-\left(\frac{\hat{\mathrm{p}}^{2}}{2\mathrm{m}}+S\right)(\chi+\chi^{*})-(\mathrm{m}c^{2}-V)\chi,
\end{equation}
and from the solution $\chi$ of the latter equation, we can write
$\Psi=\left[\,\varphi\;\,\chi\,\right]^{\mathrm{T}}=\left[\,\chi^{*}\;\,\chi\,\right]^{\mathrm{T}}=\Psi_{c}$.
Certainly, it is sufficient to solve at least one of these one-component
equations because $\chi$ and $\varphi$ are algebraically related.
Thus, it can be said that Eq. (21) alone, or Eq. (22) alone, describe
this kind of 1D KFGM particle. Similarly, if we now use the Majorana
condition $\Psi=-\Psi_{c}$, i.e., $\varphi=-\chi^{*}$ (or $\chi=-\varphi^{*}$),
we can write the following wave equation for this 1D KFGM particle:
\begin{equation}
\hat{\mathrm{E}}\,\varphi=\left(\frac{\hat{\mathrm{p}}^{2}}{2\mathrm{m}}+S\right)(\varphi-\varphi^{*})+(\mathrm{m}c^{2}+V)\varphi,
\end{equation}
in which case the respective wavefunction $\Psi$ is given by $\Psi=\left[\,\varphi\;\,\chi\,\right]^{\mathrm{T}}=\left[\,\varphi\;\,-\varphi^{*}\,\right]^{\mathrm{T}}=-\Psi_{c}$.
Alternatively, we can write the following wave equation: 
\begin{equation}
\hat{\mathrm{E}}\,\chi=-\left(\frac{\hat{\mathrm{p}}^{2}}{2\mathrm{m}}+S\right)(\chi-\chi^{*})-(\mathrm{m}c^{2}-V)\chi,
\end{equation}
in which case the respective wavefunction $\Psi$ is given by $\Psi=\left[\,\varphi\;\,\chi\,\right]^{\mathrm{T}}=\left[\,-\chi^{*}\;\,\chi\,\right]^{\mathrm{T}}=-\Psi_{c}$.
Certainly, any of these last two equations models the 1D KFGM particle
which is characterized by the condition $\Psi=-\Psi_{c}$. Incidentally,
we note that none of these last four equations for the 1D KFGM particle
is of the form $(\hat{\mathrm{E}}-\hat{\mathrm{H}})\phi=0$.

\subsection{A 1D KFGM particle in an interval}

\noindent It is important to mention that up to this point, we have
not imposed any particular or specific condition on the Hamiltonian
$\hat{\mathrm{h}}$, for example, we have not yet imposed on $\hat{\mathrm{h}}$
the condition of formal pseudohermiticity, i.e., $\hat{\mathrm{h}}_{\mathrm{adj}}\equiv\hat{\tau}_{3}\,\hat{\mathrm{h}}^{\dagger}\,\hat{\tau}_{3}=\hat{\mathrm{h}}$
(the symbol $^{\mathrm{\dagger}}$ denotes the usual Hermitian conjugate
of a matrix and an operator). Effectively, the generalized Hermitian
conjugate, or the formal generalized adjoint of $\hat{\mathrm{h}}$,
that is, $\hat{\mathrm{h}}_{\mathrm{adj}}$, is defined as 
\begin{equation}
\hat{\mathrm{h}}_{\mathrm{adj}}\equiv\hat{\tau}_{3}\,\hat{\mathrm{h}}^{\dagger}\,\hat{\tau}_{3},
\end{equation}
and is given by
\begin{equation}
\hat{\mathrm{h}}_{\mathrm{adj}}=\frac{\hat{\mathrm{p}}^{2}}{2\mathrm{m}}(\hat{\tau}_{3}+\mathrm{i}\hat{\tau}_{2})+\mathrm{m}c^{2}\hat{\tau}_{3}+V^{*}\,\hat{1}_{2}+S\,(\hat{\tau}_{3}+\mathrm{i}\hat{\tau}_{2}).
\end{equation}
In fact, we introduce the following pseudo inner product \cite{RefA}:
\begin{equation}
\langle\langle\Psi,\Phi\rangle\rangle\equiv\int_{\Omega}\mathrm{d}x\,\Psi^{\dagger}\hat{\tau}_{3}\Phi,
\end{equation}
where $\Omega=[a,b]$ (an interval), and $\Psi=\left[\,\varphi\;\,\chi\,\right]^{\mathrm{T}}$
and $\Phi=\left[\,\zeta\;\,\xi\,\right]^{\mathrm{T}}$, we can verify
that the definition given in Eq. (25) can also be formally written
as follows:
\begin{equation}
\langle\langle\hat{\mathrm{h}}_{\mathrm{adj}}\Psi,\Phi\rangle\rangle=\langle\langle\Psi,\hat{\mathrm{h}}\Phi\rangle\rangle.
\end{equation}
Specifically, this last relation requires only the definitions of
$\hat{\mathrm{h}}_{\mathrm{adj}}$ (see Eq. (25)) and the scalar product
given in Eq. (27). Actually, the relation in Eq. (28) defines the
generalized Hermitian conjugate, or the generalized adjoint $\hat{\mathrm{h}}_{\mathrm{adj}}$
on an indefinite inner product space. Then, the Hamiltonian operator
in Eq. (17) is formally pseudo-Hermitian or formally generalized Hermitian
because it satisfies the following formal relation:
\begin{equation}
\hat{\mathrm{h}}_{\mathrm{adj}}=\mathrm{\hat{h}}.
\end{equation}
Consequently, the potential $V$ must be real (compare $\mathrm{\hat{h}}$
in Eq. (17) with $\hat{\mathrm{h}}_{\mathrm{adj}}$ in Eq. (26)),
i.e., $V=V^{*}$, and because we want to characterize a 1D KFGM particle
(then $V=-V^{*}$), $V$ must be zero. Indeed, a real or complex electric
interaction does not seem to affect a particle that is strictly neutral. 

Thus, the 1D FV wave equation that describes a 1D KFGM particle is
given by Eq. (17) with $V=0$ and $S\in\mathbb{R}$, namely,
\begin{equation}
\hat{\mathrm{E}}\,\hat{1}_{2}\Psi=\mathrm{\hat{h}}\Psi=\left[\,\frac{\hat{\mathrm{p}}^{2}}{2\mathrm{m}}(\hat{\tau}_{3}+\mathrm{i}\hat{\tau}_{2})+\mathrm{m}c^{2}\hat{\tau}_{3}+S\,(\hat{\tau}_{3}+\mathrm{i}\hat{\tau}_{2})\,\right]\Psi,
\end{equation}
where $\Psi=\left[\,\varphi\;\,\varphi^{*}\,\right]^{\mathrm{T}}$
or $\Psi=\left[\,\chi^{*}\;\,\chi\,\right]^{\mathrm{T}}$ (with $\Psi=\Psi_{c}$
being the Majorana condition); equivalently, Eq. (21), or Eq. (22),
with $V=0$ and $S\in\mathbb{R}$, also describes this 1D KFGM particle.
Equally, Eq. (30) describes a 1D KFGM particle with $\Psi=\left[\,\varphi\;\,-\varphi^{*}\,\right]^{\mathrm{T}}$
or $\Psi=\left[\,-\chi^{*}\;\,\chi\,\right]^{\mathrm{T}}$ (where
$\Psi=-\Psi_{c}$ is the Majorana condition); equivalently, Eq. (23),
or Eq. (24), with $V=0$ and $S\in\mathbb{R}$, also describes this
1D KFGM particle. 

By virtue of two integrations by parts, the Hamiltonian operator $\hat{\mathrm{h}}$
in Eq. (30) and its formal generalized adjoint $\hat{\mathrm{h}}_{\mathrm{adj}}$
(which acts as the operator $\mathrm{\hat{h}}$) satisfy the following
relation:
\begin{equation}
\langle\langle\hat{\mathrm{h}}_{\mathrm{adj}}\Psi,\Phi\rangle\rangle=\langle\langle\Psi,\hat{\mathrm{h}}\Phi\rangle\rangle-\frac{\hbar^{2}}{2\mathrm{m}}\,\frac{1}{2}\left.\left[\,\left((\hat{\tau}_{3}+\mathrm{i}\hat{\tau}_{2})\Psi_{x}\right)^{\dagger}(\hat{\tau}_{3}+\mathrm{i}\hat{\tau}_{2})\Phi-\left((\hat{\tau}_{3}+\mathrm{i}\hat{\tau}_{2})\Psi\right)^{\dagger}(\hat{\tau}_{3}+\mathrm{i}\hat{\tau}_{2})\Phi_{x}\,\right]\right|_{a}^{b},
\end{equation}
where $\left.\left[\, g\,\right]\right|_{a}^{b}\equiv g(b,t)-g(a,t)$,
and $\Psi_{x}\equiv\partial\Psi/\partial x$, etc. It is worth mentioning
that the latter relation is also valid if the Hamiltonians $\mathrm{\hat{h}}$
and $\hat{\mathrm{h}}_{\mathrm{adj}}$ contain, in addition to a real
scalar potential $S$, a real electric potential $V$. Certainly,
in this specific case, we are not using the Majorana condition, i.e.,
we are not considering a 1D KFGM particle inside an interval (because
the latter case requires that $V=0$), but only a 1D KFG particle
in an interval \cite{RefV}. 

Then, if the boundary conditions imposed on $\Psi$ and $\Phi$ at
the ends of interval $\Omega$ lead to the vanishing of the boundary
term in Eq. (31), the Hamiltonian $\hat{\mathrm{h}}$, formally satisfying
Eq. (29) (i.e., $\hat{\mathrm{h}}_{\mathrm{adj}}=\hat{\mathrm{h}}$),
is effectively pseudo-Hermitian (or generalized Hermitian). The most
general family of boundary conditions leading to the cancellation
of the boundary term in Eq. (31) was obtained in Ref. \cite{RefV}.
For all the boundary conditions inside this family, $\hat{\mathrm{h}}$
is a pseudo-Hermitian operator, but it is also a pseudo self-adjoint
operator \cite{RefV}, that is, $\hat{\mathrm{h}}$ satisfies the
relation
\begin{equation}
\langle\langle\hat{\mathrm{h}}\Psi,\Phi\rangle\rangle=\langle\langle\Psi,\hat{\mathrm{h}}\Phi\rangle\rangle.
\end{equation}
Thus, the functions belonging to the domains of $\hat{\mathrm{h}}$
and $\hat{\mathrm{h}}_{\mathrm{adj}}$ obey the same boundary conditions,
and $\hat{\mathrm{h}}_{\mathrm{adj}}=\hat{\mathrm{h}}$ (in this case,
the latter is not just a formal equality). Then, the most general
set of pseudo self-adjoint boundary conditions for the Hamiltonian
$\hat{\mathrm{h}}$ is given by (for clarity, we omit the variable
$t$ in the boundary conditions hereafter) 
\begin{equation}
\left[\begin{array}{c}
(\hat{\tau}_{3}+\mathrm{i}\hat{\tau}_{2})(\Psi-\mathrm{i}\lambda\Psi_{x})(b)\\
(\hat{\tau}_{3}+\mathrm{i}\hat{\tau}_{2})(\Psi+\mathrm{i}\lambda\Psi_{x})(a)
\end{array}\right]=\hat{\mathrm{U}}_{(4\times4)}\left[\begin{array}{c}
(\hat{\tau}_{3}+\mathrm{i}\hat{\tau}_{2})(\Psi+\mathrm{i}\lambda\Psi_{x})(b)\\
(\hat{\tau}_{3}+\mathrm{i}\hat{\tau}_{2})(\Psi-\mathrm{i}\lambda\Psi_{x})(a)
\end{array}\right],
\end{equation}
where $\hat{\mathrm{U}}_{(4\times4)}$ is a $4\times4$ unitary matrix
which can be written as follows:
\begin{equation}
\hat{\mathrm{U}}_{(4\times4)}=\hat{\mathrm{S}}^{\dagger}\left[\begin{array}{cc}
\hat{\mathrm{U}}_{(2\times2)} & \hat{0}\\
\hat{0} & \hat{\mathrm{U}}_{(2\times2)}
\end{array}\right]\hat{\mathrm{S}},
\end{equation}
with 
\begin{equation}
\hat{\mathrm{S}}=\left[\begin{array}{cccc}
1 & 0 & 0 & 0\\
0 & 0 & 1 & 0\\
0 & -1 & 0 & 0\\
0 & 0 & 0 & -1
\end{array}\right]
\end{equation}
($\hat{\mathrm{S}}^{\dagger}=\hat{\mathrm{S}}^{-1}$), and as we will
see immediately, $\hat{\mathrm{U}}_{(2\times2)}$ is a $2\times2$
unitary matrix that depends on three real parameters.

The boundary term in Eq. (31) can also be written in terms of the
one-component wavefunctions corresponding to the two-component column
vectors $\Psi=\left[\,\varphi\;\,\chi\,\right]^{\mathrm{T}}$ and
$\Phi=\left[\,\zeta\;\,\xi\,\right]^{\mathrm{T}}$, namely, $\psi=\varphi+\chi$
and $\phi=\zeta+\xi$ (see Eq. (3)), evaluated at the endpoints of
the interval $\Omega$. Certainly, the relation given in Eq. (31)
can be written as follows: 
\begin{equation}
\langle\langle\hat{\mathrm{h}}_{\mathrm{adj}}\Psi,\Phi\rangle\rangle=\langle\langle\Psi,\hat{\mathrm{h}}\Phi\rangle\rangle-\frac{\hbar^{2}}{2\mathrm{m}}\left.\left[\,\psi_{x}^{*}\,\phi\,-\psi^{*}\phi_{x}\,\right]\right|_{a}^{b}.
\end{equation}
The boundary term in the latter relation is proportional to the total
derivative with respect to time of the pseudo scalar product given
in Eq. (27), where $\Psi$ and $\Phi$ are solutions of the 1D FV
wave equation, that is,
\begin{equation}
-\frac{\hbar^{2}}{2\mathrm{m}}\left.\left[\,\psi_{x}^{*}\,\phi\,-\psi^{*}\phi_{x}\,\right]\right|_{a}^{b}=\frac{\hbar}{\mathrm{i}}\,\frac{\mathrm{d}}{\mathrm{d}t}\langle\langle\Psi,\Phi\rangle\rangle.
\end{equation}
It is also important to note that the pseudo inner product $\langle\langle\Psi,\Phi\rangle\rangle$
in Eq. (37) does not depend on the Lorentz scalar potential $S$,
but it depends on the electric potential $V$ (here, we have $V=0$).
However, its time derivative is independent of the two potentials
(provided they are real valued). Then, because $\hat{\mathrm{h}}$
is a pseudo self-adjoint operator, the boundary term in Eq. (36) is
zero. The most general family of pseudo self-adjoint boundary conditions
for $\hat{\mathrm{h}}$, which is similar for the operator $\hat{\mathrm{h}}_{\mathrm{adj}}$
(because $\hat{\mathrm{h}}_{\mathrm{adj}}$ is equal to $\hat{\mathrm{h}}$,
i.e., their actions and domains are equal) and consistent with the
cancellation of the boundary term in Eq. (36), is given by 
\begin{equation}
\left[\begin{array}{c}
\psi(b)-\mathrm{i}\lambda\psi_{x}(b)\\
\psi(a)+\mathrm{i}\lambda\psi_{x}(a)
\end{array}\right]=\hat{\mathrm{U}}_{(2\times2)}\left[\begin{array}{c}
\psi(b)+\mathrm{i}\lambda\psi_{x}(b)\\
\psi(a)-\mathrm{i}\lambda\psi_{x}(a)
\end{array}\right],
\end{equation}
where $\hat{\mathrm{U}}_{(2\times2)}$ is precisely the $2\times2$
unitary matrix that appears in Eq. (34) \cite{RefV}. Let us note
that if $\hat{\mathrm{U}}_{(2\times2)}$ is known, by using Eqs. (34)
and (35), the matrix $\hat{\mathrm{U}}_{(4\times4)}$ can be calculated
immediately. Thus, the pseudo inner product $\langle\langle\Psi,\Phi\rangle\rangle$
in Eq. (37) is constant, i.e., for all the corresponding solutions
$\psi$ and $\phi$ of the standard 1D KFG wave equation that satisfy
any of the boundary conditions included in Eq. (38) (see Eqs. (36)
and (37)). 

Naturally, the standard 1D KFG wave equation also describes a 1D KFGM
particle when $V=0$ and $S\in\mathbb{R}$, namely,
\begin{equation}
\left[\,-\hbar^{2}\frac{\partial^{2}}{\partial t^{2}}+\hbar^{2}c^{2}\frac{\partial^{2}}{\partial x^{2}}-(\mathrm{m}c^{2})^{2}-2\,\mathrm{m}c^{2}S\,\right]\psi=0
\end{equation}
(see Eq. (2)), where $\psi=\psi^{*}$ ($\Psi=\Psi_{c}$ is the Majorana
condition). That is, the solutions of Eq. (39) must be written as
real. Furthermore, we can impose the Majorana condition $\Psi=-\Psi_{c}$,
and therefore, $\psi=-\psi^{*}$; however, these purely imaginary
solutions can and should also be written as real. Consequently, when
the Majorana condition is given by $\Psi=\Psi_{c}$, it follows that
both $\psi$ and $\psi^{*}$ satisfy the general boundary condition
in Eq. (38), and when the Majorana condition is $\Psi=-\Psi_{c}$,
it follows that both $\psi$ and $-\psi^{*}$ satisfy it. In these
two cases, the matrix $\hat{\mathrm{U}}_{(2\times2)}$ satisfies the
following condition:
\begin{equation}
\hat{\mathrm{U}}_{(2\times2)}^{\mathrm{T}}=\hat{\mathrm{U}}_{(2\times2)},
\end{equation}
that is, $\hat{\mathrm{U}}_{(2\times2)}$ must additionally be a (complex)
symmetric matrix. If we choose the following general expression for
$\hat{\mathrm{U}}_{(2\times2)}$: 
\begin{equation}
\hat{\mathrm{U}}_{(2\times2)}=\mathrm{e}^{\mathrm{i}\,\mu}\left[\begin{array}{cc}
\mathrm{m}_{0}-\mathrm{i}\,\mathrm{m}_{3} & -\mathrm{m}_{2}-\mathrm{i}\,\mathrm{m}_{1}\\
\mathrm{m}_{2}-\mathrm{i}\,\mathrm{m}_{1} & \mathrm{m}_{0}+\mathrm{i}\,\mathrm{m}_{3}
\end{array}\right],
\end{equation}
where $\mu\in[0,\pi)$, and the real quantities $\mathrm{m}_{k}$
($k=0,1,2,3$) satisfy $(\mathrm{m}_{0})^{2}+(\mathrm{m}_{1})^{2}+(\mathrm{m}_{2})^{2}+(\mathrm{m}_{3})^{3}=1$,
and impose on it the condition given in Eq. (40), we obtain the result
$\mathrm{m}_{2}=0$. Thus, the most general set of pseudo self-adjoint
boundary conditions for a 1D KFGM particle, or for the ultimately
real solutions of the 1D KFG wave equation in Eq. (39), is given by
\begin{equation}
\left[\begin{array}{c}
\psi(b)-\mathrm{i}\lambda\psi_{x}(b)\\
\psi(a)+\mathrm{i}\lambda\psi_{x}(a)
\end{array}\right]=\mathrm{e}^{\mathrm{i}\,\mu}\left[\begin{array}{cc}
\mathrm{m}_{0}-\mathrm{i}\,\mathrm{m}_{3} & -\mathrm{i}\,\mathrm{m}_{1}\\
-\mathrm{i}\,\mathrm{m}_{1} & \mathrm{m}_{0}+\mathrm{i}\,\mathrm{m}_{3}
\end{array}\right]\left[\begin{array}{c}
\psi(b)+\mathrm{i}\lambda\psi_{x}(b)\\
\psi(a)-\mathrm{i}\lambda\psi_{x}(a)
\end{array}\right],
\end{equation}
and depends on three real parameters. Indeed, the square matrix $\hat{\mathrm{U}}_{(2\times2)}$
in Eq. (42) is the one that determines the matrix $\hat{\mathrm{U}}_{(4\times4)}$
that appears in the more general set of pseudo self-adjoint boundary
conditions for the 1D KFGM particle (see Eqs. (33) and (34)). The
latter general family of boundary conditions is for the solutions
of the 1D FV wave equation in Eq. (30), which can also describe a
1D KFGM particle. Incidentally, because the matrix $\hat{\mathrm{S}}$
in Eq. (35) is real, one has that $\hat{\mathrm{S}}^{\dagger}=\hat{\mathrm{S}}^{\mathrm{T}}$,
but in addition, $\hat{\mathrm{U}}_{(2\times2)}$ satisfies Eq. (40),
consequently, the matrix $\hat{\mathrm{U}}_{(4\times4)}$ in Eq. (34)
is also a (complex) symmetric matrix, i.e., $\hat{\mathrm{U}}_{(4\times4)}^{\mathrm{T}}=\hat{\mathrm{U}}_{(4\times4)}$.
Actually, if $\hat{\mathrm{U}}_{(2\times2)}$ is the matrix given
in Eq. (41) with $\mathrm{m}_{2}=0$, then $\hat{\mathrm{U}}_{(4\times4)}$
is given by
\begin{equation}
\hat{\mathrm{U}}_{(4\times4)}=\mathrm{e}^{\mathrm{i}\,\mu}\left[\begin{array}{cc}
(\mathrm{m}_{0}-\mathrm{i}\,\mathrm{m}_{3})\hat{1}_{2} & -\mathrm{i}\,\mathrm{m}_{1}\hat{1}_{2}\\
-\mathrm{i}\,\mathrm{m}_{1}\hat{1}_{2} & (\mathrm{m}_{0}+\mathrm{i}\,\mathrm{m}_{3})\hat{1}_{2}
\end{array}\right].
\end{equation}

Some of the boundary conditions included in the general set of pseudo
self-adjoint boundary conditions for a 1D particle KFGM given in Eq.
(42) are the following: (i) $\psi(a)=\psi(b)=0$ ($\mathrm{m}_{0}=+1$,
$\mathrm{m}_{1}=\mathrm{m}_{3}=0$ and $\mu=\pi$); (ii) $\psi_{x}(a)=\psi_{x}(b)=0$
($\mathrm{m}_{0}=+1$, $\mathrm{m}_{1}=\mathrm{m}_{3}=0$ and $\mu=0$);
(iii) $\psi(a)-\lambda\psi_{x}(a)=0$ and $\psi(b)+\lambda\psi_{x}(b)=0$
($\mathrm{m}_{0}=+1$, $\mathrm{m}_{1}=\mathrm{m}_{3}=0$ and $\mu=\pi/2$);
(iv) $\psi(a)=\psi(b)$ and $\psi_{x}(a)=\psi_{x}(b)$ ($\mathrm{m}_{0}=\mathrm{m}_{3}=0$,
$\mathrm{m}_{1}=+1$ and $\mu=\pi/2$); (v) $\psi(a)=-\psi(b)$ and
$\psi_{x}(a)=-\psi_{x}(b)$ ($\mathrm{m}_{0}=\mathrm{m}_{3}=0$, $\mathrm{m}_{1}=-1$
and $\mu=\pi/2$). In this short list, we have distinguished boundary
conditions: (i) is the Dirichlet boundary condition, (ii) is the Neumann
condition, (iii) is a kind of Robin boundary condition, (iv) is the
periodic condition and (v) is the antiperiodic condition. As discussed
above, a 1D KFGM particle supports only those boundary conditions
arising from the unitary matrix $\hat{\mathrm{U}}_{(2\times2)}$ in
Eq. (41) with $\mathrm{m}_{2}=0$. For example, some boundary conditions
that are not suitable for a 1D KFGM particle but are suitable for
a 1D KFG particle ($\mathrm{m}_{2}\neq0$) are the following: (vi)
$\psi(a)=\pm\mathrm{i}\psi(b)$ and $\psi_{x}(a)=\pm\mathrm{i}\psi_{x}(b)$
($\mathrm{m}_{0}=\mathrm{m}_{1}=\mathrm{m}_{3}=0$, $\mathrm{m}_{2}=\pm1$
and $\mu=\pi/2$); (vii) $\psi(a)=\pm\mathrm{i}\lambda\psi_{x}(b)$
and $\psi(b)=\pm\mathrm{i}\lambda\psi_{x}(a)$ ($\mathrm{m}_{0}=\mathrm{m}_{1}=\mathrm{m}_{3}=0$,
$\mathrm{m}_{2}=\pm1$ and $\mu=0$). 

Due to the Majorana condition, the components of the wave function
$\Psi$ in Eq. (30) are not independent. When this condition is given
by $\Psi=\Psi_{c}$, then $\chi=\varphi^{*}$, and therefore, $\psi=\varphi+\chi=\varphi+\varphi^{*}=2\,\mathrm{Re}(\varphi)$
and $\psi_{x}=2\,(\mathrm{Re}(\varphi))_{x}$. Thus, the equation
in (21) with $V=0$ and $S\in\mathbb{R}$, namely, 
\begin{equation}
\hat{\mathrm{E}}\,\varphi=\left(\frac{\hat{\mathrm{p}}^{2}}{2\mathrm{m}}+S\right)(\varphi+\varphi^{*})+\mathrm{m}c^{2}\varphi,
\end{equation}
describes a 1D KFGM particle (the one for which $\Psi=\Psi_{c}$),
and its solutions must satisfy any of the boundary conditions included
in the general set of boundary conditions given in Eq. (42) but written
only in terms of $\varphi$, namely, 
\begin{equation}
\left[\begin{array}{c}
(\mathrm{Re}(\varphi))(b)-\mathrm{i}\lambda(\mathrm{Re}(\varphi))_{x}(b)\\
(\mathrm{Re}(\varphi))(a)+\mathrm{i}\lambda(\mathrm{Re}(\varphi))_{x}(a)
\end{array}\right]=\mathrm{e}^{\mathrm{i}\,\mu}\left[\begin{array}{cc}
\mathrm{m}_{0}-\mathrm{i}\,\mathrm{m}_{3} & -\mathrm{i}\,\mathrm{m}_{1}\\
-\mathrm{i}\,\mathrm{m}_{1} & \mathrm{m}_{0}+\mathrm{i}\,\mathrm{m}_{3}
\end{array}\right]\left[\begin{array}{c}
(\mathrm{Re}(\varphi))(b)+\mathrm{i}\lambda(\mathrm{Re}(\varphi))_{x}(b)\\
(\mathrm{Re}(\varphi))(a)-\mathrm{i}\lambda(\mathrm{Re}(\varphi))_{x}(a)
\end{array}\right].
\end{equation}
Let us note the simultaneous presence of $\varphi$ and $\varphi^{*}$
in Eq. (44); however, from this equation, it follows that the real
part of $\varphi$ satisfies the 1D KFG equation, namely,
\begin{equation}
\left[\,-\hbar^{2}\frac{\partial^{2}}{\partial t^{2}}+\hbar^{2}c^{2}\frac{\partial^{2}}{\partial x^{2}}-(\mathrm{m}c^{2})^{2}-2\,\mathrm{m}c^{2}S\,\right]\mathrm{Re}(\varphi)=0,
\end{equation}
and the imaginary part of $\varphi$ can be obtained by taking the
time derivative of the real part, namely,
\begin{equation}
\mathrm{Im}(\varphi)=\frac{\hbar}{\mathrm{m}c^{2}}\,\frac{\partial}{\partial t}\,\mathrm{Re}(\varphi).
\end{equation}
Clearly, if the scalar potential depends explicitly on time, the imaginary
part of $\varphi$ does not satisfy Eq. (46)  (see Appendix A). Finally,
the solutions of Eq. (44) are simply given by $\varphi=\mathrm{Re}(\varphi)+\mathrm{i}\,\mathrm{Im}(\varphi)$
(and the component $\chi$ of $\Psi$ is obtained from the Majorana
condition, i.e., $\chi=\varphi^{*}$). As discussed above, in this
same case ($\Psi=\Psi_{c}$), the equation in (22) can alternatively
be used with $V=0$ and $S\in\mathbb{R}$, namely,
\begin{equation}
\hat{\mathrm{E}}\,\chi=-\left(\frac{\hat{\mathrm{p}}^{2}}{2\mathrm{m}}+S\right)(\chi+\chi^{*})-\mathrm{m}c^{2}\chi.
\end{equation}
However, in addition, in Eq. (42), the relations $\psi=\varphi+\chi=\chi^{*}+\chi=2\,\mathrm{Re}(\chi)$
and $\psi_{x}=2\,(\mathrm{Re}(\chi))_{x}$ must be used, namely,  
\begin{equation}
\left[\begin{array}{c}
(\mathrm{Re}(\chi))(b)-\mathrm{i}\lambda(\mathrm{Re}(\chi))_{x}(b)\\
(\mathrm{Re}(\chi))(a)+\mathrm{i}\lambda(\mathrm{Re}(\chi))_{x}(a)
\end{array}\right]=\mathrm{e}^{\mathrm{i}\,\mu}\left[\begin{array}{cc}
\mathrm{m}_{0}-\mathrm{i}\,\mathrm{m}_{3} & -\mathrm{i}\,\mathrm{m}_{1}\\
-\mathrm{i}\,\mathrm{m}_{1} & \mathrm{m}_{0}+\mathrm{i}\,\mathrm{m}_{3}
\end{array}\right]\left[\begin{array}{c}
(\mathrm{Re}(\chi))(b)+\mathrm{i}\lambda(\mathrm{Re}(\chi))_{x}(b)\\
(\mathrm{Re}(\chi))(a)-\mathrm{i}\lambda(\mathrm{Re}(\chi))_{x}(a)
\end{array}\right].
\end{equation}
In this case, it can be shown that $\mathrm{Re}(\chi)$ satisfies
the same Eq. (46) and any of the boundary conditions in Eq. (49).
The imaginary part of $\chi$ is obtained from the following relation:
\begin{equation}
\mathrm{Im}(\chi)=-\frac{\hbar}{\mathrm{m}c^{2}}\,\frac{\partial}{\partial t}\,\mathrm{Re}(\chi).
\end{equation}
Thus, the solutions of Eq. (48) are simply given by $\chi=\mathrm{Re}(\chi)+\mathrm{i}\,\mathrm{Im}(\chi)$
(and the component $\varphi$ of $\Psi$ is obtained from the Majorana
condition, i.e., $\varphi=\chi^{*}$).

Similarly, when the Majorana condition is given by $\Psi=-\Psi_{c}$,
$\chi=-\varphi^{*}$, and therefore, $\psi=\varphi+\chi=\varphi-\varphi^{*}=2\mathrm{i}\,\mathrm{Im}(\varphi)$
and $\psi_{x}=2\mathrm{i}\,(\mathrm{Im}(\varphi))_{x}$. Thus, the
equation in (23) with $V=0$ and $S\in\mathbb{R}$, namely,
\begin{equation}
\hat{\mathrm{E}}\,\varphi=\left(\frac{\hat{\mathrm{p}}^{2}}{2\mathrm{m}}+S\right)(\varphi-\varphi^{*})+\mathrm{m}c^{2}\varphi,
\end{equation}
characterizes a 1D KFGM particle (the one for which $\Psi=-\Psi_{c}$)
and its solutions must satisfy some of the boundary conditions given
in Eq. (42), but the latter equation written in terms of $\varphi$
only, specifically, 
\begin{equation}
\left[\begin{array}{c}
(\mathrm{Im}(\varphi))(b)-\mathrm{i}\lambda(\mathrm{Im}(\varphi))_{x}(b)\\
(\mathrm{Im}(\varphi))(a)+\mathrm{i}\lambda(\mathrm{Im}(\varphi))_{x}(a)
\end{array}\right]=\mathrm{e}^{\mathrm{i}\,\mu}\left[\begin{array}{cc}
\mathrm{m}_{0}-\mathrm{i}\,\mathrm{m}_{3} & -\mathrm{i}\,\mathrm{m}_{1}\\
-\mathrm{i}\,\mathrm{m}_{1} & \mathrm{m}_{0}+\mathrm{i}\,\mathrm{m}_{3}
\end{array}\right]\left[\begin{array}{c}
(\mathrm{Im}(\varphi))(b)+\mathrm{i}\lambda(\mathrm{Im}(\varphi))_{x}(b)\\
(\mathrm{Im}(\varphi))(a)-\mathrm{i}\lambda(\mathrm{Im}(\varphi))_{x}(a)
\end{array}\right].
\end{equation}
From Eq. (51), the imaginary part of $\varphi$ satisfies the 1D KFG
equation, as expected, that is, 
\begin{equation}
\left[\,-\hbar^{2}\frac{\partial^{2}}{\partial t^{2}}+\hbar^{2}c^{2}\frac{\partial^{2}}{\partial x^{2}}-(\mathrm{m}c^{2})^{2}-2\,\mathrm{m}c^{2}S\,\right]\mathrm{Im}(\varphi)=0,
\end{equation}
with the boundary conditions given in Eq. (52). The real part of $\varphi$
is obtained from the relation 
\begin{equation}
\mathrm{Re}(\varphi)=-\frac{\hbar}{\mathrm{m}c^{2}}\,\frac{\partial}{\partial t}\,\mathrm{Im}(\varphi).
\end{equation}
Finally, $\varphi=\mathrm{Re}(\varphi)+\mathrm{i}\,\mathrm{Im}(\varphi)$
is the solution of Eq. (51) (and the component $\chi$ of $\Psi$
is obtained from the Majorana condition, i.e., $\chi=-\varphi^{*}$).
Equivalently (because $\Psi=-\Psi_{c}$ is still valid), the equation
in (24) can also be used with $V=0$ and $S\in\mathbb{R}$, namely,
\begin{equation}
\hat{\mathrm{E}}\,\chi=-\left(\frac{\hat{\mathrm{p}}^{2}}{2\mathrm{m}}+S\right)(\chi-\chi^{*})-\mathrm{m}c^{2}\chi.
\end{equation}
However, now, in Eq. (42), the relations $\psi=\varphi+\chi=-\chi^{*}+\chi=2\mathrm{i}\,\mathrm{Im}(\chi)$
and $\psi_{x}=2\mathrm{i}\,(\mathrm{Im}(\chi))_{x}$ must be used.
Then, we can write
\begin{equation}
\left[\begin{array}{c}
(\mathrm{Im}(\chi))(b)-\mathrm{i}\lambda(\mathrm{Im}(\chi))_{x}(b)\\
(\mathrm{Im}(\chi))(a)+\mathrm{i}\lambda(\mathrm{Im}(\chi))_{x}(a)
\end{array}\right]=\mathrm{e}^{\mathrm{i}\,\mu}\left[\begin{array}{cc}
\mathrm{m}_{0}-\mathrm{i}\,\mathrm{m}_{3} & -\mathrm{i}\,\mathrm{m}_{1}\\
-\mathrm{i}\,\mathrm{m}_{1} & \mathrm{m}_{0}+\mathrm{i}\,\mathrm{m}_{3}
\end{array}\right]\left[\begin{array}{c}
(\mathrm{Im}(\chi))(b)+\mathrm{i}\lambda(\mathrm{Im}(\chi))_{x}(b)\\
(\mathrm{Im}(\chi))(a)-\mathrm{i}\lambda(\mathrm{Im}(\chi))_{x}(a)
\end{array}\right].
\end{equation}
The latter boundary conditions are for the solutions of the wave equation
in (55), which can also describe the 1D KFGM particle for which $\Psi=-\Psi_{c}$.
In this case, it can be shown that $\mathrm{Im}(\chi)$ also satisfies
Eq. (53) and any of the boundary conditions in Eq. (56) (which are
certainly the same boundary conditions that are satisfied by $\mathrm{Im}(\varphi)$),
and the real part of $\chi$ is obtained from the relation
\begin{equation}
\mathrm{Re}(\chi)=\frac{\hbar}{\mathrm{m}c^{2}}\,\frac{\partial}{\partial t}\,\mathrm{Im}(\chi).
\end{equation}
Finally, the solutions of Eq. (55) are obtained from $\chi=\mathrm{Re}(\chi)+\mathrm{i}\,\mathrm{Im}(\chi)$
(and the component $\varphi$ of $\Psi$ is obtained from the Majorana
condition, i.e., $\varphi=-\chi^{*}$).

We may refer to Eqs. (44), (48), (51) and (55) as the first-order
(non-Hamiltonian) 1D Majorana equations in time for the 1D KFGM particle,
and their solutions are complex. On the other hand, all the examples
of  the boundary conditions we presented above for the wavefunction
$\psi$ have identical counterparts for $\mathrm{Re}(\varphi)$ and
$\mathrm{Re}(\chi)$, and $\mathrm{Im}(\varphi)$ and $\mathrm{Im}(\chi)$.
This is because the families of boundary conditions given in Eqs.
(45), (49), and (52), (56), are similar in form. Incidentally, each
of the first-order 1D Majorana equations leads to a second-order 1D
Majorana equation. None of the latter equations is the standard 1D
KFG equation; however, when the scalar potential does not explicitly
depend on time, each equation becomes the standard 1D KFG equation.
This is exhibited in the Appendix A. 

Last, if we set $\Psi=\Phi$ in Eq. (36), and therefore, $\psi=\phi$,
we obtain
\begin{equation}
\langle\langle\hat{\mathrm{h}}_{\mathrm{adj}}\Psi,\Psi\rangle\rangle=\langle\langle\Psi,\hat{\mathrm{h}}\Psi\rangle\rangle-\frac{\hbar}{\mathrm{i}}\left.\left[\, j\,\right]\right|_{a}^{b},
\end{equation}
where
\begin{equation}
j=j(x,t)=\frac{\mathrm{i}\hbar}{2\mathrm{m}}\left(\,\psi_{x}^{*}\,\psi-\psi^{*}\psi_{x}\,\right)=\frac{\hbar}{\mathrm{m}}\,\mathrm{Im}\left(\,\psi^{*}\psi_{x}\,\right)
\end{equation}
is the 1D KFG ``probability'' current density (although it is certainly
not correct to interpret it as a quantity representing a probability).
Similarly, because the pseudo inner product $\langle\langle\Psi,\Phi\rangle\rangle$
in Eq. (37) is independent of time, $\langle\langle\Psi,\Psi\rangle\rangle\equiv\int_{\Omega}\mathrm{d}x\,\Psi^{\dagger}\hat{\tau}_{3}\Psi=\int_{\Omega}\mathrm{d}x\,\varrho$
is also a constant quantity, where 
\begin{equation}
\varrho=\varrho(x,t)=\Psi^{\dagger}\hat{\tau}_{3}\Psi=|\varphi|^{2}-|\chi|^{2}=\frac{1}{2\mathrm{m}c^{2}}\left[\,\psi^{*}(\hat{\mathrm{E}}\psi)-(\hat{\mathrm{E}}\psi^{*})\,\psi\,\right]
\end{equation}
is the 1D KFG ``probability'' density, but we also have that $\left.\left[\, j\,\right]\right|_{a}^{b}=0$.
Therefore, $j(b,t)=j(a,t)$ (see Eq. (37)). These two quantities,
$j=j(x,t)$ and $\varrho=\varrho(x,t)$, also satisfy the continuity
equation in this situation where a real scalar potential exists, namely,
$\partial\varrho/\partial t=-\partial j/\partial x$ (certainly, the
latter equation is valid for the solutions of the 1D KFG equation
with a scalar potential, for example, the solutions $\psi$ of Eq.
(39)). Let us note that if $\varrho$ and $j$ are nonzero, the continuity
equation can be integrated in $x$, which leads to the expected result
$\tfrac{\mathrm{d}}{\mathrm{d}t}\int_{\Omega}\mathrm{d}x\,\varrho=-\left.\left[\, j\,\right]\right|_{a}^{b}$.
Then, because $\left.\left[\, j\,\right]\right|_{a}^{b}=0$, we have
that $\int_{\Omega}\mathrm{d}x\,\varrho=\mathrm{const}$.

Clearly, when the solutions of Eq. (39) are definitely real-valued
functions, the 1D KFGM current density $j$ and density $\varrho$
cease to exist. That is, when $\psi=\psi^{*}$, the expressions for
$j$ and $\varrho$ given in Eqs. (59) and (60) are identically zero;
however, when $\psi=-\psi^{*}$, the result is the same, namely, 
\begin{equation}
j=j(x,t)=0\quad\mathrm{and}\quad\varrho=\varrho(x,t)=0.
\end{equation}
Thus, ultimately, the reality condition for the wavefunction $\psi$
also automatically leads to the so-called impenetrability condition
at the extremes of the interval $\Omega$, i.e., $j(b,t)=j(a,t)=0$.
This situation contrasts with the case of the 1D Dirac wave equation
in the Majorana representation (by considering the Dirac theory as
a single-particle theory). There, the Majorana condition is given
by $\Psi_{\mathrm{D}}=\Psi_{\mathrm{D}}^{*}$, so that only real solutions
describe the 1D Dirac-Majorana particle; however, the corresponding
probability current density is not automatically zero \cite{RefYY}.
In conclusion, regardless of which boundary condition one takes from
the general set in Eq. (42), which is satisfied by a real solution
$\psi$, this solution does, by necessity, always verify the mathematical
condition of impenetrability at the walls of the interval.

In any case, it is important to remember that for a 1D KFG particle
moving in a finite interval (see Eq. (2), where $V$ and $S$ are
real-valued potentials, and therefore, their solutions $\psi$ are
always complex-valued functions), we can recognize impenetrability
boundary conditions (or confining boundary conditions) and nonconfining
boundary conditions \cite{RefV}. In general, the confining boundary
conditions satisfy $j(b,t)=j(a,t)=0$, and nonconfining boundary conditions
simply satisfy $j(b,t)=j(a,t)$ (for which it is certainly necessary
that the solutions $\psi$ are complex). However, when the solutions
of Eq. (2) are real-valued functions (which can occur when $V=0$);
that is, when the solutions of the equation describing a 1D KFGM particle
are real (see Eq. (39)), the distinction between confining and nonconfining
boundary conditions is not feasible, at least if the current density
$j$ is considered. Certainly, the characterization of these two types
of boundary conditions would require the use of some other current
density. We hope to study this issue in a forthcoming paper.

\section{Summary and conclusions}

\noindent In the first quantization, the wave equations considered
to describe a strictly neutral 1D KFGM particle are the standard 1D
KFG equation and/or the 1D FV equation, both with a real Lorentz scalar
potential plus their respective Majorana conditions. Unexpectedly,
one finds that the Majorana condition appears here in two specific
forms, say, one standard and one nonstandard. Specifically, we showed
that the imposition of the standard (nonstandard) Majorana condition
on the solutions of the 1D FV equation implies that the solutions
of the second order 1D KFG equation in the time must be real (imaginary;
however, they can also be written real, as expected). Additionally,
both Majorana conditions determine that the scalar potential must
be real. In any case, we found that the solutions of the time-dependent
1D FV equation cannot be real. We also showed that the additional
imposition of the formal pseudohermiticity condition on the Hamiltonian
that is present in this equation together with a Majorana condition
determines that the electric potential must be zero. In addition,
if we place a 1D KFGM particle in a finite interval, then the corresponding
Hamiltonian is a pseudo self-adjoint operator. As a consequence of
this property, one has a three-parameter general set of boundary conditions
for the solutions of the 1D FV equation and another for the respective
real solutions of the standard 1D KFG equation. We found that these
two general sets of boundary conditions are the same for the two types
of Majorana conditions. Because of the Majorana condition, the components
of the wavefunction for the 1D FV equation are not independent; hence,
we wrote first-order equations in time for each of these components
and obtained the general sets of pseudo self-adjoint boundary conditions
that they must obey. Incidentally, these equations do not have a Hamiltonian
form, but any of them alone can model a 1D KFGM particle (in fact,
if one of the complex components of the solution of the 1D FV equation
is known, the other component can be obtained algebraically via the
Majorana condition). Thus, we refer to these equations as the first-order
1D Majorana equations for the 1D KFGM particle. Additionally, we wrote
second-order 1D Majorana equations in time for each of the components
of the 1D FV equation. These equations become the standard 1D KFG
equation when the scalar potential does not explicitly depend on time
(see Appendix A). 

As shown in Appendix B, the nonrelativistic limit of the first-order
1D Majorana equation given in Eq. (44) yields the partial differential
equation given in Eq. (B11). The latter equation is not the Schr\"{o}dinger
equation because the term enclosed in the bracket does not have to
be zero. Having said that, if it is assumed that the solution of this
nonrelativistic equation $\varphi_{\mathrm{NR}}$ and its complex
conjugate $\varphi_{\mathrm{NR}}^{*}$ can be treated as independent
solutions, then $\varphi_{\mathrm{NR}}$ satisfies the Schr\"{o}dinger
equation and $\varphi_{\mathrm{NR}}^{*}$ satisfies the equation which
is the complex conjugate of the Schr\"{o}dinger equation. A similar situation
arises when studying the nonrelativistic limit of certain real scalar
field theories (see, for example, Refs. \cite{RefXX,RefWW,RefVV,RefUU,RefTT}).
There, the (classical) relativistic field is real, i.e., $\psi=\psi^{*}$,
but the nonrelativistic $\psi_{\mathrm{NR}}$ is complex; thus, in
that case, by taking the nonrelativistic limit, the typical ansatz
we used in Appendix B must be modified. Apropos of this, the first-order
Majorana equations we introduce here describe strictly neutral particles,
and their solutions are always complex; thus, the ansatz can be the
usual one. Finally, in each of those field theories (and only in certain
cases), the nonrelativistic Schr\"{o}dinger equation and its respective
complex conjugate equation could be obtained if the solutions of these
two equations are assumed to be independent. Incidentally, the latter
assumption has been questioned (see, for example, Ref. \cite{RefXX}).
In closing, our results can also be extended to the problem of a 1D
KFGM particle in a real line with a tiny hole at a point, for example,
at $x=0$ (i.e., $\Omega=\mathbb{R}-\{0\}$). Indeed, the general
sets of pseudo self-adjoint boundary conditions for this problem can
be obtained from those corresponding to the particle within the interval
$\Omega=[a,b]$ by identifying the ends of the interval with the two
sides of the hole, namely, $x=a\rightarrow0+$ and $x=b\rightarrow0-$. 

\section{Appendix A}

\noindent As we have seen, we can write $2+2=4$ first-order (non-Hamiltonian)
1D Majorana equations in time for the 1D KFGM particles, namely, Eqs.
(44) and (48) (by using the standard Majorana condition), and (51)
and (55) (by using the nonstandard Majorana condition). Likewise,
we can also write four second-order 1D Majorana equations in time
for these particles, i.e., four second-order equations for the components
$\varphi$ and $\chi$ of $\Psi$. In effect, applying the operator
$\hat{\mathrm{E}}$ to both sides of Eq. (44), and using the relation
$\hat{\mathrm{E}}^{*}=-\hat{\mathrm{E}}$, gives the following equation:
\[
\left[\,\hat{\mathrm{E}}^{2}-(c\,\hat{\mathrm{p}})^{2}-(\mathrm{m}c^{2})^{2}-2\,\mathrm{m}c^{2}S\,\right]\varphi=(\hat{\mathrm{E}}\, S)(\varphi+\varphi^{*}).\tag{A1}
\]
Similarly, from Eq. (48), the following equation is obtained:
\[
\left[\,\hat{\mathrm{E}}^{2}-(c\,\hat{\mathrm{p}})^{2}-(\mathrm{m}c^{2})^{2}-2\,\mathrm{m}c^{2}S\,\right]\chi=-(\hat{\mathrm{E}}\, S)(\chi+\chi^{*}).\tag{A2}
\]
These two equations correspond to the Majorana condition $\Psi=\Psi_{c}$,
that is, $\psi=\psi^{*}$. If we add Eqs. (A1) and (A2) and use the
relations given in Eqs. (3) and (4) (the latter with $V=0$), it is
confirmed that $\psi=\varphi+\chi$ satisfies the 1D KFG equation
(i.e., Eq. (39)), namely, 
\[
\left[\,\hat{\mathrm{E}}^{2}-(c\,\hat{\mathrm{p}})^{2}-(\mathrm{m}c^{2})^{2}-2\,\mathrm{m}c^{2}S\,\right]\psi=0,\tag{A3}
\]
as expected. Note that only when the scalar potential does not explicitly
depend on time, the complex components $\varphi$ and $\chi$ of $\Psi$
also satisfy this equation. If this is not the case, only the functions
$\mathrm{Re}(\varphi)$ and $\mathrm{Re}(\chi)$ can satisfy the 1D
KFG equation (see the discussion following Eq. (45) through Eq. (50)). 

Similarly, applying the operator $\hat{\mathrm{E}}$ to both sides
of Eq. (51), and using the relation $\hat{\mathrm{E}}^{*}=-\hat{\mathrm{E}}$,
gives the equation
\[
\left[\,\hat{\mathrm{E}}^{2}-(c\,\hat{\mathrm{p}})^{2}-(\mathrm{m}c^{2})^{2}-2\,\mathrm{m}c^{2}S\,\right]\varphi=(\hat{\mathrm{E}}\, S)(\varphi-\varphi^{*}).\tag{A4}
\]
In the same manner, applying $\hat{\mathrm{E}}$ to Eq. (55) gives
the following equation:
\[
\left[\,\hat{\mathrm{E}}^{2}-(c\,\hat{\mathrm{p}})^{2}-(\mathrm{m}c^{2})^{2}-2\,\mathrm{m}c^{2}S\,\right]\chi=-(\hat{\mathrm{E}}\, S)(\chi-\chi^{*}).\tag{A5}
\]
The latter two equations correspond to the Majorana condition $\Psi=-\Psi_{c}$,
that is, $\psi=-\psi^{*}$. If we add Eqs. (A4) and (A5) and use the
relations given in Eqs. (3) and (4) (the latter with $V=0$), it is
again found that $\psi=\varphi+\chi$ satisfies the 1D KFG equation
(i.e., Eq. (A3)), as expected. Clearly, if $(\hat{\mathrm{E}}\, S)=0$,
then $\varphi$ and $\chi$ also satisfy the 1D KFG equation. If $(\hat{\mathrm{E}}\, S)\neq0$,
then only the functions $\mathrm{Im}(\varphi)$ and $\mathrm{Im}(\chi)$
can satisfy this equation (see the discussion following Eq. (53) through
Eq. (57)).

\section{Appendix B}

\noindent Let us examine the nonrelativistic approximation of one
of the first-order 1D Majorana equations in time, for example, Eq.
(44). As we know, the latter equation for $\varphi\in\mathbb{C}$
is completely equivalent to Eq. (46), namely,
\[
\left[\,-\hbar^{2}\frac{\partial^{2}}{\partial t^{2}}+\hbar^{2}c^{2}\frac{\partial^{2}}{\partial x^{2}}-(\mathrm{m}c^{2})^{2}-2\,\mathrm{m}c^{2}S\,\right]\mathrm{Re}(\varphi)=0,\tag{B1}
\]
plus the relation given in Eq. (47), namely,
\[
\mathrm{Im}(\varphi)=\frac{\hbar}{\mathrm{m}c^{2}}\,\frac{\partial}{\partial t}\,\mathrm{Re}(\varphi).\tag{B2}
\]
Let us note that Eq. (B1) can also be written as follows: 
\[
\mathrm{Re}\left[\,\left(\,-\hbar^{2}\frac{\partial^{2}}{\partial t^{2}}+\hbar^{2}c^{2}\frac{\partial^{2}}{\partial x^{2}}-(\mathrm{m}c^{2})^{2}-2\,\mathrm{m}c^{2}S\,\right)\varphi\,\right]=0.\tag{B3}
\]
Now, we choose the typical ansatz that connects $\varphi$ to its
nonrelativistic approximation $\varphi_{\mathrm{NR}}$, namely,
\[
\varphi=\varphi_{\mathrm{NR}}\,\mathrm{e}^{-\mathrm{i}\frac{\mathrm{m}c^{2}}{\hbar}t},\tag{B4}
\]
and therefore, 
\[
\varphi_{t}=\left[(\varphi_{\mathrm{NR}})_{t}-\mathrm{i}\frac{\mathrm{m}c^{2}}{\hbar}\,\varphi_{\mathrm{NR}}\right]\mathrm{e}^{-\mathrm{i}\frac{\mathrm{m}c^{2}}{\hbar}t}\tag{B5}
\]
and 
\[
\varphi_{tt}=\left[(\varphi_{\mathrm{NR}})_{tt}-\mathrm{i}\frac{2\mathrm{m}c^{2}}{\hbar}\,(\varphi_{\mathrm{NR}})_{t}-\frac{(\mathrm{m}c^{2})^{2}}{\hbar^{2}}\,\varphi_{\mathrm{NR}}\right]\mathrm{e}^{-\mathrm{i}\frac{\mathrm{m}c^{2}}{\hbar}t}.\tag{B6}
\]
In the nonrelativistic approximation, we have that 
\[
\left|\,\mathrm{i}\hbar\,(\varphi_{\mathrm{NR}})_{t}\,\right|\ll\mathrm{m}c^{2}\left|\,\varphi_{\mathrm{NR}}\,\right|\;\Rightarrow\;\left|\,(\varphi_{\mathrm{NR}})_{t}\,\right|\ll\frac{\mathrm{m}c^{2}}{\hbar}\left|\,\varphi_{\mathrm{NR}}\,\right|\tag{B7}
\]
and 
\[
\left|\,\mathrm{i}\hbar\,(\varphi_{\mathrm{NR}})_{tt}\,\right|\ll\mathrm{m}c^{2}\left|\,(\varphi_{\mathrm{NR}})_{t}\,\right|\;\Rightarrow\;\left|\,(\varphi_{\mathrm{NR}})_{tt}\,\right|\ll\frac{\mathrm{m}c^{2}}{\hbar}\left|\,(\varphi_{\mathrm{NR}})_{t}\,\right|.\tag{B8}
\]
Consequently, in this regime, the relations given in Eqs. (B5) and
(B6) can be written as follows:
\[
\varphi_{t}=-\mathrm{i}\frac{\mathrm{m}c^{2}}{\hbar}\,\varphi_{\mathrm{NR}}\,\mathrm{e}^{-\mathrm{i}\frac{\mathrm{m}c^{2}}{\hbar}t}\tag{B9}
\]
and 
\[
\varphi_{tt}=\left[-\mathrm{i}\frac{2\mathrm{m}c^{2}}{\hbar}\,(\varphi_{\mathrm{NR}})_{t}-\frac{(\mathrm{m}c^{2})^{2}}{\hbar^{2}}\,\varphi_{\mathrm{NR}}\right]\mathrm{e}^{-\mathrm{i}\frac{\mathrm{m}c^{2}}{\hbar}t}.\tag{B10}
\]
Substituting the latter expression into Eq. (B3), we obtain the following
result:
\[
\mathrm{Re}\left[\,\mathrm{e}^{-\mathrm{i}\frac{\mathrm{m}c^{2}}{\hbar}t}\left(\,-\hat{\mathrm{E}}+\frac{\hat{\mathrm{p}}^{2}}{2\mathrm{m}}+S\,\right)\varphi_{\mathrm{NR}}\,\right]=0.\tag{B11}
\]
Clearly, this is not the Schr\"{o}dinger equation with the scalar interaction,
i.e., $\varphi_{\mathrm{NR}}$ in Eq. (B11) does not necessarily obey
this equation. 

Similarly, the relation that gives the imaginary part of $\varphi$
(Eq. (B2)) can also be written as follows:
\[
\mathrm{Im}(\varphi)=\mathrm{Re}\left(\frac{\hbar}{\mathrm{m}c^{2}}\,\varphi_{t}\right).\tag{B12}
\]
Substituting Eqs. (B4) and (B9) into Eq. (B12), we obtain the result
\[
\mathrm{Im}\left(\varphi_{\mathrm{NR}}\,\mathrm{e}^{-\mathrm{i}\frac{\mathrm{m}c^{2}}{\hbar}t}\right)=\mathrm{Re}\left(-\mathrm{i}\,\varphi_{\mathrm{NR}}\,\mathrm{e}^{-\mathrm{i}\frac{\mathrm{m}c^{2}}{\hbar}t}\right),\tag{B13}
\]
which is always true because $\mathrm{Im}(z)=\mathrm{Re}(-\mathrm{i}z)$,
for all $z\in\mathbb{C}$. Thus, nothing new is obtained from Eq.
(B2) and the nonrelativistic limit of Eq. (44) reduces to Eq. (B11).
Finally, $\varphi_{\mathrm{NR}}$ is obtained from Eq. (B11), $\varphi$
is given in Eq. (B4) and $\chi=\varphi^{*}$. 

\section*{Conflicts of interest}

\noindent The author declares no conflicts of interest.

\section*{Acknowledgment}

\noindent The author wishes to thank the reviewers for their comments
and suggestions.

\end{document}